\documentclass[aps,pre,reprint,groupedaddress,longbibliography,floatfix]{revtex4-1}

\usepackage{amsmath}
\usepackage{amssymb}
\usepackage{graphicx,psfrag}
\usepackage{subfig}
\usepackage{verbatim}
\usepackage{color}

\begin{document}


\title{Integer Lattice Gas with a sampling collision operator for the fluctuating Diffusion Equation}


\author{Noah Seekins}
\email{noah.seekins@ndsu.edu}
\author{Alexander J. Wagner}
\email[]{alexander.wagner@ndsu.edu}
\homepage[]{www.ndsu.edu/pubweb/$\sim$carswagn}
\affiliation{Department of Physics, North Dakota State University, Fargo, North Dakota 58108, USA}


\date{\today}

\begin{abstract}
  We developed an integer lattice gas method for the fluctuating diffusion equation. Such a method is unconditionally stable and able to recover the Poisson distribution for the microscopic densities.  A key advance for integer lattice gases introduced in this paper is a new sampling collision operator that replaces particle collisions with sampling from an equilibrium distribution. This can increase the efficiency of our integer lattice gas by several orders of magnitude.
\end{abstract}

\keywords{Lattice Gas, Lattice Boltzmann, Monte Carlo, Fluctuatons}

\maketitle


\section{Introduction}
The idea of the integer lattice gas developed by Blommel \textit{et al.}~\cite{blommel2018integer} extends traditional Boolean lattice gases \cite{FHP,doolen1991lattice}, that allow only one particle per lattice link to instead allow for an integer number of particles. This cures problems with the local density dependent advection pre-factor that breaks Galilean invariance in those standard lattice gas model. It also includes ideal gas fluctuations in a consistent manner, and can therefore represent fluctuations for low densities.

Historically lattice Boltzmann methods were derived from lattice gas models~\cite{FHP,wolf2004lattice}. The exclusion principle that allowed for one particle per lattice velocity only caused the equilibrium distribution to be of the Fermi-Dirac rather than the Boltzmann form. This equilibrium distribution was the reason, that the hydrodynamic limit of these lattice gases contained non Galilean invariant terms. The original lattice Boltzmann methods were exactly equivalent to these lattice gas methods~\cite{higuera1989boltzmann}.  Qian and d'Humieres modified the lattice Boltzmann collision operator to break the link to the underlying lattice gas model~\cite{qian1992lattice}, which enabled them to recover the Galilean invariance of the lattice gas hydrodynamics. Blommel and Wagner~\cite{blommel2018integer} were able to show that one can define a set of integer lattice gas models which will have a corresponding entropic lattice Boltzmann method as its Boltzmann averaged limit. This equivalence showed that the integer lattice gases share the improved level of Galilean invariance with their lattice Boltzmann counterparts, a significant improvement over Boolean lattice gases. The recovery of a  Galilean invariant (in the lattice Boltzmann sense, since no lattice based method can be fully invariant) lattice gas was striking, but the advance was mostly of theoretical interest, since the fundamental two particle lattice gas collision operator was not computationally competitive with the corresponding lattice Boltzmann method.

In this paper we develop a sampling collision operator for the lattice gas method, and we show that with such a collision operator the lattice gas can be competitive with the corresponding lattice Boltzmann method. Such a sampling approach was previously suggested by Boghosian~\cite{boghosian1997integer}, but in that case it was considered impractical to construct such a sampling collision operator. We show here that by focusing on a diffusive system leading to a diffusive lattice Boltzmann method~\cite{wolf1995lattice}, it is indeed possible to construct an efficient sampling method for our integer lattice gas.

In section II we introduce the lattice gas, and in section III we derive its Boltzmann average. Section IV is dedicated to validating the fluctuating and dynamic properties of the method for a few test cases. Section V is dedicated to the analysis of the improved  computational efficiency of the method.

\section{The Monte Carlo Lattice Gas algorithm}
A lattice gas consists of an underlying lattice and a set of lattice velocities $v_i$ as well as occupation numbers $n_i(x,t)$ indicating the number of particles at lattice point $x$ at time $t$ associated with lattice velocity $v_i$. The distance between the nearest neighbors of the lattice sites constitutes the lattice spacing $\Delta x$. Time increments in discrete time steps $\Delta t$. The velocities $v_i$ are defined such that the lattice displacements $v_i\Delta t$ are lattice vectors, i.e. they connect lattice sites with each other. In particular if $x$ is a lattice site so is $x+v_i\Delta t$. Typically the lattice velocities are the same for each lattice site and restricted in number such that the lattice displacements are restricted to a (small) neighborhood. In general we will refer to the number of lattice velocities as $V$.

When we say that for each lattice velocity $v_i$ there is an associated integer occupation number $n_i(x,t)$, what we mean is $n_i(x,t)$ that the number of particles at lattice site $x$ at time $t$ that came from lattice site $x-v_i$ in the previous time step. This allows us to define a local number of particles at each lattice site
\begin{equation}
    N(x,t)=\sum_i n_i(x,t).
    \label{eqn:N}
\end{equation}
The time-evolution of the lattice occupation numbers $n_i$ can be written as
\begin{equation}
  n_i(x+v_i\Delta t,t+\Delta t)=n_i(x,t)+\Xi_i,
\end{equation}
where $\Xi_i$ is referred to as the collision operator. This collision operator will be a stochastic operator in general that obeys the local conservation laws. In our case the local number of particles $N(x,t)$ will be left unchanged by the collision. In the integer lattice gas developed by Blommel \textit{et al.}~\cite{blommel2018integer} this collision operator was constructed as the net effect of many two-particle collisions that conserved mass and momentum. This allowed for an implementation that recovered independent Poisson distributed fluctuations of the occupation numbers as well as an equilibrium distribution derivable from a maximum entropy consideration. The relaxation towards equilibrium was analytically derived, even including a non-linear relaxation regime. Despite these attractive features, the method did not constitute a viable replacement for fluctuating lattice Boltzmann methods as the collision operator required an execution time that, for larger numbers of collisions, would lead to times significantly larger than the equivalent lattice Boltzmann execution time. This prevented the integer lattice gas from becoming the method of choice for most practical applications.

To alleviate this drawback we consider here a collision operator that performs all collisions in one step by directly sampling from the local equilibrium distribution. We expect that this will increase the performance of integer lattice gas method. While we believe that this is possible in the general case, this paper focuses on the simpler case that only has mass (i.e. no momentum) conservation. The hydrodynamic limit of such a method recovers the diffusion equation. We leave the extension to hydrodynamic lattice gasses to a future research project. We therefore aim to derive the lattice gas that is equivalent to the fluctuating lattice Boltzmann method for the diffusion equation~\cite{wagner2016fluctuating}.

As in the case of the integer gas of Blommel \textit{et al.}~\cite{blommel2018integer} we are imposing the weights $w_i$ for the occupation numbers in local equilibrium. The collision of Blommel \textit{et al.} involved two particle collisions such that the total number of particles as well as the momentum (although not the energy) in each collision was conserved. In the case of our diffusive lattice gas momentum is not conserved. This represents a significant simplification since only the number of particles needs to be conserved and we  now can consider single-particle collisions, which can be thought of physically as collisions of the particles with a matrix or a solvent that is not explicitly modeled. 

Collision rules equivalent to those of Blommel's integer lattice gas consist of selecting a particle at random. Since particles are not individually labeled in a lattice gas,  this corresponds to picking a particle associated with velocity $v_{s}$ with probability $p_{s}=n_{s}/\sum_j n_j$. We impose a local equilibrium distribution for a number of particles $N$ at the lattice site, as a local ensemble average. This is an average over many collisions, but with a fixed number of particles, i.e. without exchanging particles with neighbouring cells. This local equilibrium is then given by
\begin{equation}
    f_i^{0}(N)=\langle n_i\rangle^0 = N w_i,
    \label{eqn:f0}
\end{equation}
where $\langle \cdots \rangle^0$ implies an equilibrium average over occupation numbers $n_i$ under local collisions, but without an exchange of particle with neighbouring sites. The $w_i$ are weights with 
\begin{equation}
    1 = \sum_{i=1}^V w_i,
\end{equation}
equivalent to the weights used in lattice Boltzmann~\cite{qian1992lattice} which are derived as a discretization of a Maxwell-Boltzmann distribution onto the lattice velocities $v_i$.

In local equilibrium we require that detailed balance, \textit{i.e.} in equilibrium the transition from state A to state B is just as likely as the reverse transition, is obeyed. This can be formally written as a constraint on the general transition probabilities $P_{i\rightarrow j}$ :
\begin{equation}
  N w_i P_{i\rightarrow j} = N w_j P_{j\rightarrow i},
  \label{semidetailed}
\end{equation}
This tells us that the transition probabilities $P_{i\rightarrow j}$ for $i\neq j$ obey
\begin{equation}
  P_{i\rightarrow j} \propto \min\left(1,\frac{w_j}{w_i}\right).
  \label{prob}
\end{equation}
This only defines the transition probabilities up to a pre-factor, so we define
\begin{equation}
    P_{i\rightarrow j} = \left\{
    \begin{array}{cc}
        \lambda_{ij} \min\left(1,\frac{w_j}{w_i}\right) & \mbox{ for }i\neq j \\
        1-\sum_{i,j (i\neq j)} P_{i\rightarrow j} &\mbox{ for }i=j
    \end{array}\right.
\end{equation}
where $\lambda_{ij}=\lambda_{ji}$ according to Eq.~(\ref{semidetailed}). For the most efficient simulations  $\lambda_{ij}$ should be as large as possible to ensure the best possible acceptance rate for the collision. 
This becomes particularly simple, and maximally efficient, if we choose
\begin{equation}
    \lambda_{ij}=\max(w_i,w_j),
\end{equation}
and we get 
\begin{equation}
    P_{i\rightarrow j}=w_j,
\end{equation}
which clearly fulfills the detailed balance Eq.~(\ref{semidetailed}).
Practically this amounts to selecting a particle at random with  velocity $v_s$ and reassigning a new velocity $v_t$ with probability $w_t$. 

The effect of this collision on the occupation number $n_i$ can be written as a random variable
\begin{equation}
  \vartheta^c_j = \delta_{t,j}-\delta_{s,j},
\end{equation}
where the Kroneker delta $\delta_{i,j}$ is one for $i=j$ and zero otherwise. 

The collision operator can then be written as a sum of $C$ such simple collision operators
\begin{equation}
  \Xi_i^C=\sum_{c=1}^C \vartheta^c_i.
  \label{eqn:XiC}
\end{equation}
This fully defines an integer lattice gas for the fluctuating diffusion equation analogous to the integer lattice gas for hydrodynamic systems introduced by Blommel~\cite{blommel2018integer}. The disadvantage of these methods is that the time required for the collision operator scales linearly with the number of  collisions. The number of collisions scales with the numbers of particles $C\propto N$, and particularly for higher densities this approach will become slow. The key idea of this paper is to replace the collision operator of Eq. (\ref{eqn:XiC}) with a one that simply samples the post-collision distribution from an appropriately chosen distribution and thereby to perform all collisions in a single step.

In general for a system with $Q$ discrete velocities and $K$ conserved quantities this relates  to the daunting problem of sampling from a distribution that lives on $Q-K$ dimensional manifold in a $Q$ dimensional space~\cite{boghosian1997integer}. For this comparatively simple system, however, the problem becomes tractable.  

For many collisions we can obtain a unique local equilibrium probability for the set of occupation numbers $\{n_i\}$. Since $N=\sum_i n_i$ of Eq.~(\ref{eqn:N}) is the number of particles at the lattice site (which is fixed for the local collisions), the mean of these local equilibrium distributions is given by $N w_i$.  The probability for a set of occupation numbers $\{n_i\}$ is given by the multinomial distribution
\begin{equation}
  P(\{n_i\}) = \left\{ \begin{array}{ll} N! \prod_{i=1}^V \frac{w_i^{n_i}}{n_i!} &\mbox{ if } \sum_i n_i=N,\\
  \mbox{ }\\
    0 & \mbox{ otherwise.} \end{array}\right. .
  \label{Equil}
\end{equation}
This can be seen by realizing that this is entirely equivalent to the standard combinatorial problem of the cumulative occurrence of $N$ trials with $V$ outcomes with respective probabilities $w_i$. In this textbook example the probability of having $\{n_i\}$ occurrences of events $i$ is then given by Eq. (\ref{Equil}). This is often used as the example when the multinomial distribution is introcuded~\cite{Guichard}.

First we focus on the case for $\Xi^C_i$ in the limit of $C\rightarrow \infty$, i.e. the case where we sample from the equilibrium distribution of Eq.~(\ref{Equil}).

Sampling directly out of the multinomial distribution is implemented by some packages like the GNU scientific library (GSL)~\cite{GNU}.
It allows us to sample a set of $V$ random numbers with the probability given by Eq. (\ref{Equil}). We can then sample the redistribution of $N$ particles onto $V$ boxes with probability $w_i$ through
\begin{equation}
    \left(\begin{array}{c}\hat{n}_0\\\vdots\\\hat{n}_V\end{array}\right) =\left(\begin{array}{c} (X_{w_0,\cdots,w_V}^N)_0\\\vdots\\(X_{w_0,\cdots,w_V}^N)_V\end{array}\right)
    \label{eqn:Xmulti}
\end{equation}
where $(X_{w_0,\cdots,w_V}^N)_i$ is the $i^{th}$ component of a multinomial sample. Note that all this ensures that
\begin{equation}
    \sum_i \hat{n}_i = N
    \label{eqn:Ncons}
\end{equation}
\textit{i.e.} that the collision does conserve the number of particles.

Alternatively we can do this sequentially by picking new occupation numbers from a binomial distribution as follows: let $X_p^N$ be a binomially distributed random number with
\begin{equation}
  P\left(X_p^N=n\right) = \left(\begin{array}{c} N\\n\end{array}\right) p^n (1-p)^{N-n}.
\end{equation}
We will need such binomially distributed numbers later even if we use multinomially distributed random numbers from Eq. (\ref{eqn:Xmulti}).
To numerically obtain these binomially distributed random numbers one can use libraries like the GNU scientific library~\cite{GNU}. Since we expect the drawing of these random numbers to be a key factor in the overall performance of our algorithm we also developed an algorithm in-house. This algorithm is described in appendix \ref{code}, and we found that it could be significantly faster than the GSL algorithm for a range of about $N\in[10,1000]$ average particles per lattice site. 

The collision operator is then defined by the following binomial sampling algorithm. Out of the available $N$ particles we pick an occupation number associated to $v_0$ such that each particle has a probability $w_0$ to be assigned to $n_0$ given by
\begin{equation}
  \hat{n}_0 = X_{w_0}^N,
  \label{eqn:X0}
\end{equation}
leaving us with $\tilde{N}_1=N-n_0$ particles. We now pick $n_1$ particles out of the remaining $\tilde{N}_1$ particles. Because there are now fewer states available the probability of a particle to be assigned to this occupation number $n_1$, however, has increased to $\tilde{w}_1=w_1/(1-w_0)$. We then pick
\begin{equation}
  \hat{n}_1 = X_{\tilde{w}_1}^{\tilde{N}_1}.
  \label{eqn:X1}
\end{equation}
For the remaining occupation numbers we can define the remaining particles as
\begin{equation}
  \tilde{N}_i=\tilde{N}_{i-1}  -n_{i-1}
  \label{eqn:Xi}
\end{equation}
and the normalized probability as
\begin{equation}
  \tilde{w}_i= \frac{w_i}{1-\sum_{j=1}^{i-1} w_i}.
\end{equation}
We sample the remaining densities as
\begin{equation}
  \hat{n}_i = X_{\tilde{w}_i}^{\tilde{N}_{i-1}}.
  \label{eqnni}
\end{equation}
Note that because $\sum_i w_i=1$ the probability associated with the last weight $\tilde{w}_V=1$, \textit{i.e.} the remaining particles will be assigned to the last occupation number $n_V$ with probability 1. Therefore, this algorithm ensures local conservation of particles and Eqn. (\ref{eqn:Ncons}) is again fulfilled. This is of course to be expected since this implementation is just a specific recipe to generating a multinomially distributed random number and is therefore mathematically equivalent to (\ref{eqn:Xmulti}). It will be seen below, however, that either algorithm can be more efficient, depending on what context they are used in.

In this case for $C\rightarrow \infty$ the sampling collision operator is then given by
\begin{equation}
  \Xi_i^{C\to\infty} = \hat{n}_i-n_i.
  \label{fullsampling}
\end{equation}

For a finite number of collisions $C$ we need to consider the fraction of uncollided particles. The original algorithm of Blommel et al. \cite{blommel2018integer} used a fixed number of collisions, but there is no reason that there should be a fixed number of collisions at each lattice site at each time-step.  Instead, we are envisaging here a fixed collision probability for each particle during the time-step $\Delta t$. This is both more realistic on physical grounds and easier to implement as a sampling algorithm.

If each particle has the  probability $\omega$  to be collided  during the current time-step,  and  there  are  $N$  particles at a lattice site,  then  on average $N\omega$ collisions will occur. We can go a step further and find the probability distribution for the number of collided particles as
\begin{equation}
    P(N^c) =\left(  \begin{array}{c}N\\N^c\end{array}\right) \omega^{N^c} (1-\omega)^{N-N^c}.
    \label{eqn:Nselect}
\end{equation}
To reproduce a finite number of collisions with the sampling collision operator we then take the number of particles $N^c$ that will undergo a collision and remove  them at random from the occupation numbers, again using a bionmialy distributed random number. We have to remove a binomially distributed random number of particles from each of the occupation numbers. The number of particles associated with velocity $v_i$ that are undergoing a collision are then selected through
\begin{equation}
  n^\omega_i = X^{n_i}_\omega
  \label{eqn:Xomega}
\end{equation}
and we perform the sampling algorithm as above, except that we only distribute $N^c=\sum_i n^c_i$ particles. If we denote the redistributed particles from the equivalent equation to Eq~(\ref{eqnni}) as $\hat{n}^\omega_i$ then the collision operator becomes
\begin{equation}
  \Xi_i^\omega = \hat{n}^\omega_i-n^\omega_i
  \label{eqn:Xiomega}
\end{equation}
and for $\omega=1$ we recover the full sampling collision operator of Eq.~(\ref{fullsampling}). This then defines an integer lattice gas for and arbitrary number of collisions. 
  
\section{Boltzmann average of the lattice gas}
To compare the lattice gas results to the lattice Boltzmann method we will now examine an non-equilibrium ensemble average of the lattice gas evolution equation. We define the particle probability densities as 
\begin{equation}
  f_i(x,t) = \langle n_i(x,t) \rangle^{neq}
\end{equation}
where $\langle \cdots\rangle^{neq}$ implies a non-equilibrium average over an ensemble of microscopic realizations leading to the same macroscopic state. We define the density as
\begin{equation}
\rho(x,t) = \langle N(x,t)\rangle^{neq}.
\end{equation}
The evolution equation for the average particle densities is then
\begin{equation}
  f_i(x+v_i,t+1) = f_i(x,t)+\Omega_i.
  \label{MCLB}
\end{equation}
The Boltzmann collision operator can be obtained as an averaged lattice gas collision operator
\begin{align}
  \Omega_i^\omega &= \langle\; \Xi_i^\omega\rangle^{neq}\nonumber\\
  &= \langle \hat{n}^\omega_i - n^\omega_i \rangle^{neq}\nonumber\\
  &= \omega (f^0_i - f_i),
  \label{eqn:LBcoll}
\end{align}
where $f^0$ is the local equilibrium distribution function defined in Eq. (\ref{eqn:f0}).
This is exactly the form of the standard BGK lattice Boltzmann collision operator \cite{wagner2016fluctuating}. Note that because of the particular simplicity of the diffusive case this is a pure BGK collision operator, in contrast to additional non-linear contributions Blommel \textit{et al.}~\cite{blommel2018integer} found for the hydrodynamic case.

In the hydrodynamic limit, the density  obeys a diffusion equation
\begin{equation}
  \partial_t \rho = \nabla [D \nabla (\rho \theta)],
  \label{diffeq}
\end{equation}
with
\begin{align}
  D &= \left(\frac{1}{\omega}-\frac{1}{2}\right) \label{Ddef},\\
  \theta & = \frac{1}{d}\sum_i w_i v_i^2.
\end{align}
where $d$ is the number of spatial dimensions.

\section{Verification of the method}
So far we have focused on the Boltzmann average of our lattice gas. The original reason we were interested in the lattice gas were its fluctuating properties. The fluctuations in an ideal gas have been discussed by Landau \cite[\S 114]{landau1969statistical}, where it is shown that for a classical Boltzmann gas the number density in sub-volumes are Poisson distributed. The argument here is trivially extended to lattice gases showing that (for large lattices) each density $n_i$ should be Poisson distributed:
\begin{equation}
P(n_i)=\frac{exp(-f_i^{eq})(f_i^{eq})^{n_i}}{n_i!},
\label{poisson}
\end{equation}
where 
\begin{equation}
    f^{eq} = \langle n_i \rangle, 
\end{equation}
is the global equilibrium distribution, corresponding to a full equilibrium average of the system. In our case this is given by $w_i$ times the average number of particles per lattice cell.

\begin{figure}
    \centering
    \includegraphics[width=\columnwidth]{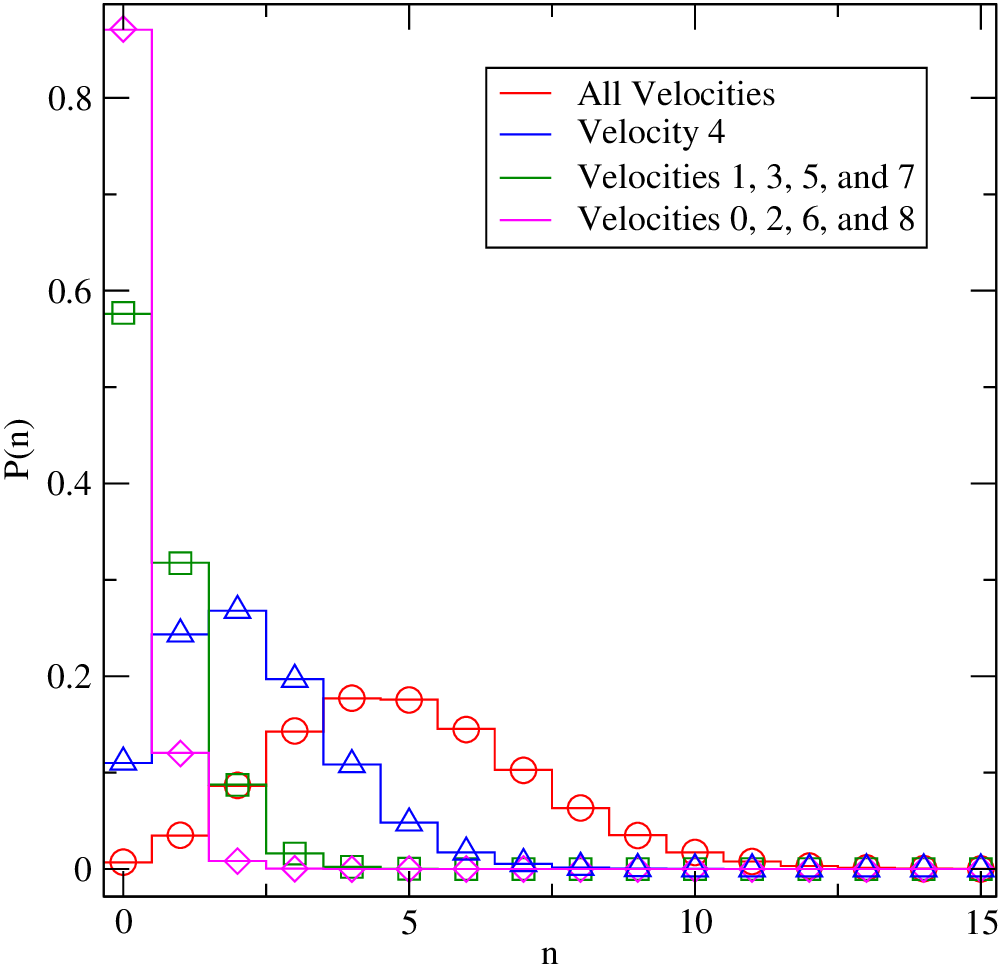}
    \caption{The $n_i$ for the integer lattice gas are Poisson distributed. This example is for an average density of $N=10$.}
    \label{fig:Poisson}
\end{figure}

For explicit examples of the verification of the method we consider a square lattice and lattice displacements consisting of all displacements of $\pm 1$ lattice sites in each dimension. We use a two dimensional system, often referred to as a D2Q9 model, where D2 refers to the fact that the system has 2 dimensions and Q9 indicates the number of lattice velocities associated with each lattice site \textit{i.e.}~$V=9$ in out notation above.

We show in Figure \ref{fig:Poisson} that we indeed recover Poisson distributions for $n_i$ as well as $\rho$ in the integer lattice gas method. In order to do this, we run a simulation initialized in a sine wave with an average of 10 particles per lattice an a 32x32 lattice grid. We then run the simulation for 10000 timesteps to allow it to relax to approximate equilibrium, and then take data points of the numbers of particles in each velocity at each lattice for 1000 timesteps, average this data, and normalize it. This measured Poisson distribution lacks some inaccuracies that resulted from continuous densities for the fluctuating lattice Boltzmann method presented in \cite{wagner2016fluctuating}.

\begin{figure}
    \centering
    \includegraphics[width=\columnwidth]{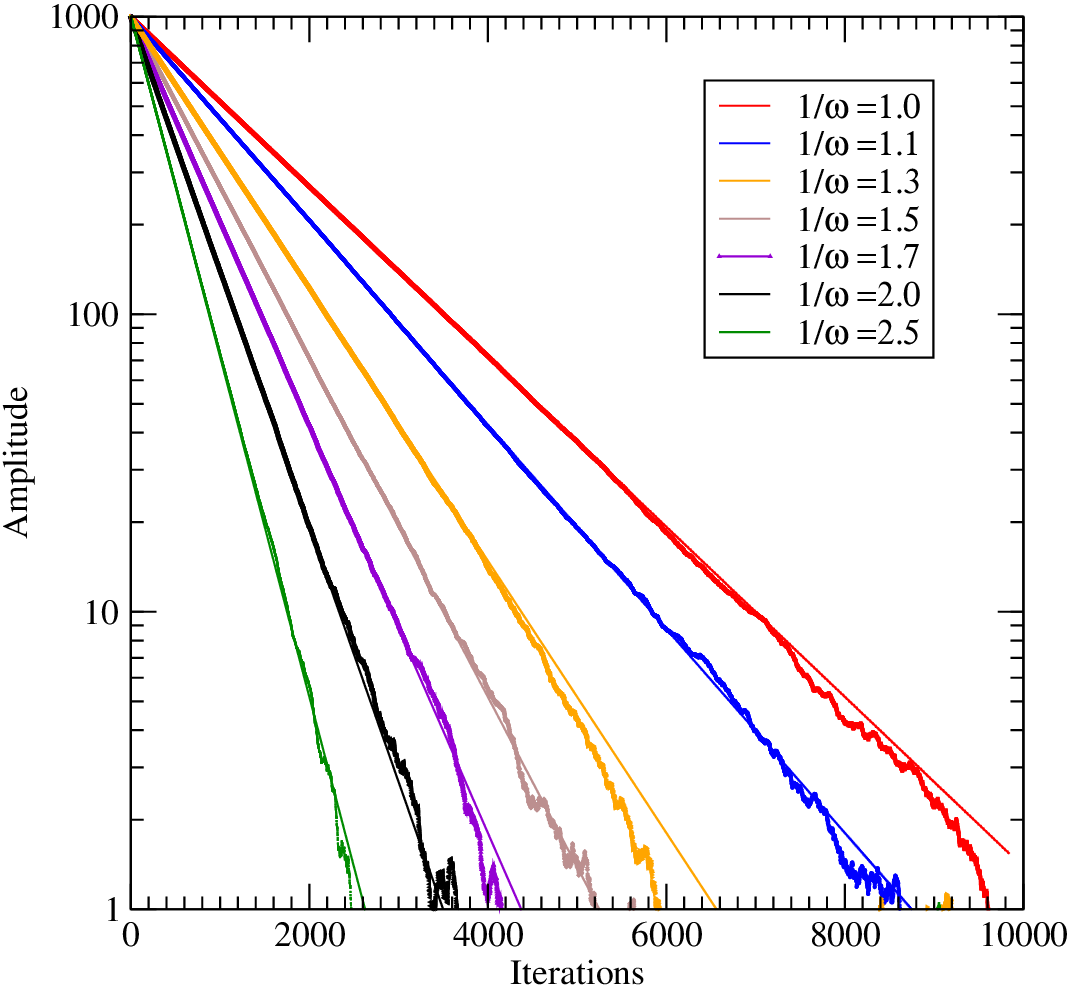}
    \caption{Amplitudes of decaying sine wave density profiles for different relaxation times, compared to theoretical predictions using a D2Q9 system of 32 by 32 lattice points with an average of 1000 particles per lattice.}
    \label{fig:amplitudes}
\end{figure}

To verify that the lattice gas does indeed have Eq. (\ref{diffeq}) as its hydrodynamic limit we look at an example problem that has an analytical solution. We initialize the simulation with a density profile in the shape of a $\sin()$ function:
\begin{equation}
    \rho(x,y,0)=N^{av}\left[1+\sin\left(\frac{2\pi x}{L_x}\right)\right]
\end{equation}
where $N^{av}$ is the average number of particles. The evolution of Eq.~(\ref{diffeq}) with this initial condition has the analytical solution 
\begin{align}
    \rho(x,y,t)=& N^{av}\left[1+\sin\left(\frac{2\pi x}{L_x}\right)\exp\left(-\frac{4\pi^2 D t}{L_x^2}\right)\right]\\
    =& N^{av} + A^{th}(t) \sin\left(\frac{2\pi x}{L_x}\right).
\end{align}
This defines the decay amplitude 
\begin{equation}
    A^\mathrm{th}(t) = N^{av}\exp\left(-\frac{4\pi^2 D t}{L_x^2}\right).
    \label{Ath}
\end{equation}
The mathematics are essentially identical to that presented in Blommel \textit{et al.} for a decaying shear wave. 
To implement an initial profile of a sine wave in a lattice gas we have the difficulty that the density is not an integer, and a lattice gas needs to include fluctuations that are already averaged away in Eq. (\ref{diffeq}). We therefore need to start with an initial density profile that is sampled from a sinusoidal probability distribution which is Poisson distributed. In particular we need
\begin{equation}
    P(\rho) = N^{av} \left[1+\sin\left(\frac{2\pi x}{L_x}\right)\right].
    \label{eqn:Prho}
\end{equation}
We achieve this by picking the occupation numbers $n_i(x,0)$ as Poisson distributed random numbers with expectation value $w_i P(\rho)$. This procedure allows us to initialize non-integer valued initial distributions and also implements the full equilibrium fluctuations in the initial configuration of the simulation.

We then the run our simulation and record the densities at each timestep.  We utilizing the method presented in Blommel \textit{et al.} to extract the amplitude from the data:
\begin{equation}
    A^\mathrm{LG}(t) = \frac{\sum_{x=1}^{L_x}\sum_{y=1}^{L_y} \sin\left(\frac{2\pi x}{L_x}\right) N(x,y,t)}{ L_y\sum_x \sin^2\left(\frac{2\pi x}{L_x}\right)}.
\end{equation}
We then compare this measured amplitude to the theoretical result of Eq. (\ref{Ath}) in Figure \ref{fig:amplitudes} and find excellent agreement. The simulation was performed on a 32x32 D2Q9 lattice system for various values of $\tau$, each lattice point having an average of 1000 particles. Over several timesteps it can be seen that these decays match the theory, up until the region where the amplitude is so small that the averaging becomes insufficient and the fluctuations take over. This result validates our prediction that we can tune the diffusion constant by allowing for only partial collisions given by Eq.  (\ref{Ddef}). An outstanding, but rather interesting, question is whether lattice gases can also implement over-relaxation, as is often done in lattice Boltzmann methods. Preliminary results suggest that this is possible, at least for larger numbers of particles per cell. A detailed discussion of this subject, however, is outside the scope of this paper.

\section{Computational efficiency}
In the previous section we have shown that the novel sampling collision operator will give results that are essentially equivalent to the direct single particle collision approach to integer lattice gases introduced by Blommel \textit{et al.}\cite{blommel2018integer}. A key ingredient in the algorithm is the sampling procedure that allows us to obtain binomially distributed random numbers in Eqs. (\ref{eqn:X0})--(\ref{eqn:Xi}) as well as Eq. (\ref{eqn:Xomega}). There exist open source sampling algorithms, and we implemented both a sequential multinomial sampling algorithm and a direct multinomial sampling algorithm using the GSL library \cite{gough2009gnu}. We further improved the performance of our algorithm within a certain range by writing our own sampling algorithm, which we refer to as the Lookup Table method below. Details of this algorithm are given in Appendix \ref{code}. 

Since we derived the sampling collision operator to improve the execution speed of the integer lattice gas method, we performed benchmarking tests of different versions of our algorithm: the base comparison is to the Collision method, \textit{i.e.} the algorithm based given by Eqn. (\ref{eqn:XiC}).

We compare the timing of this baseline method to the algorithm using the sampling collision operator either with full collision of Eq. (\ref{fullsampling}) or with partial collisions of Eq. (\ref{eqn:Xiomega}). Both algorithms either employ the GSL or Lookup Table method to generate the binomially distributed random numbers.

In addition, we compare the results to the fluctuating lattice Boltzmann method developed by Wagner \textit{et al.} \cite{wagner2016fluctuating}. There are two versions of this executable: One compiled with standard flags, and a second one using the -O3 optimization flag, which significantly improved the algorithm's runtime. Optimization flags had significant less effect on the lattice gas implementation. We summarize the different methods in Table \ref{tableMethod}. The source code for these methods is available on Github at \cite{SeekinsDiffMCLG2021Git}.

\begin{table}[]
    \begin{tabular}{l|l|l}
         Method & Collision Operator & Comment\\
         \hline
         Collision & Eq. (\ref{eqn:XiC}) & Random Collisions\\
         GSL       & Eq. (\ref{eqn:Xiomega})  & GSL Sampling\\
         GSL Multinomial & Eq. (\ref{eqn:Xiomega})  & GSL Mult. Sampling\\
         Lookup Table & Eq. (\ref{eqn:Xiomega}) & Lookup Table Sampling\\
         LB & Eq. (\ref{eqn:LBcoll})+noise & Fluctuating LB\\
         Optimized LB & Eq. (\ref{eqn:LBcoll}) & LB with the -O3 flag\\
    \end{tabular}
    \caption{Summary of the different algorithms compared in the timing benchmarks.}
    \label{tableMethod}
\end{table}

\begin{figure}
    \centering
    \includegraphics[width=\columnwidth]{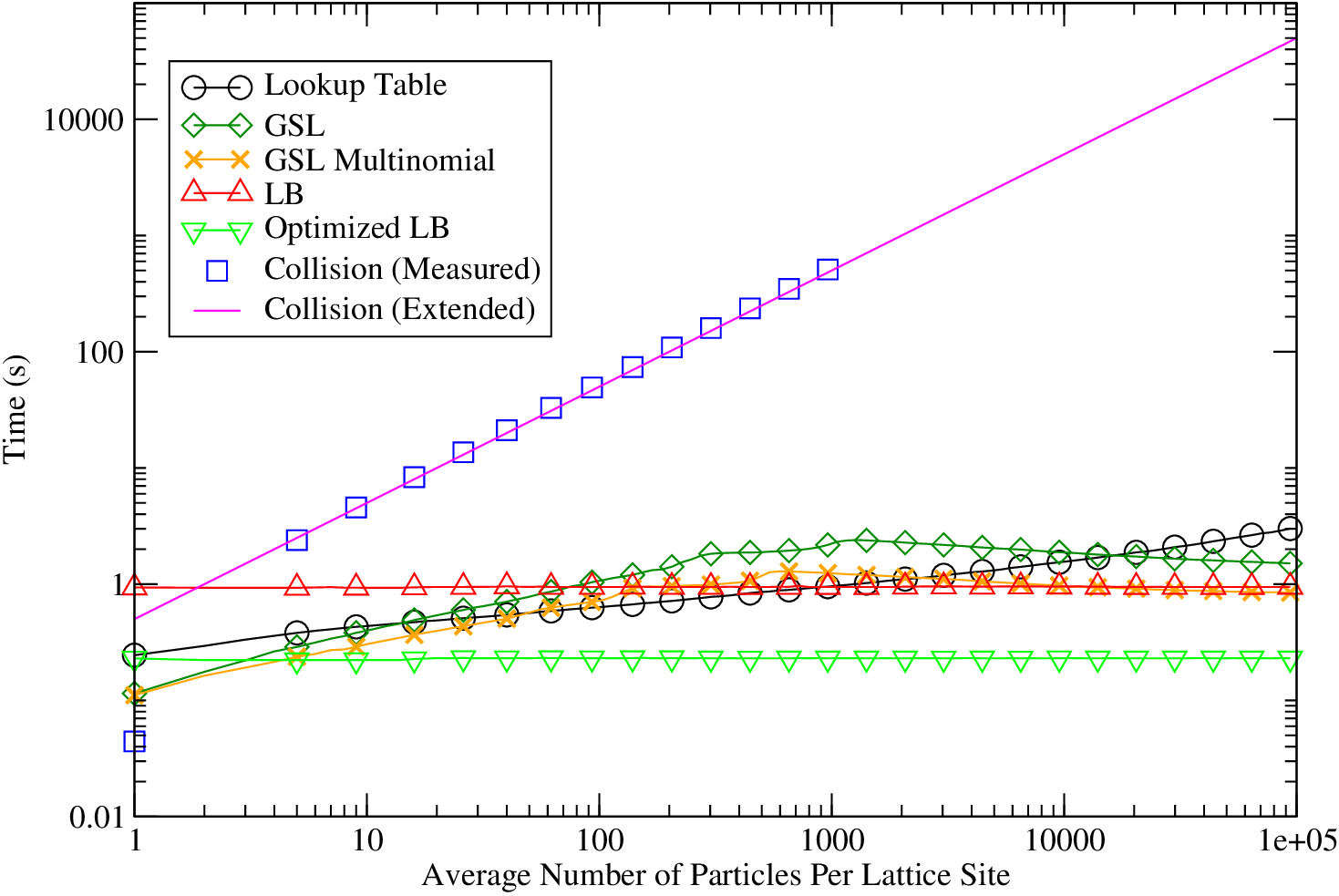}
    \caption{Execution time of the different versions of the sampling algorithm compared to the original integer lattice gas approach and two lattice Boltzmann runs graphed on a log-log scale. Each method was tested using a 32 x 32 lattice grid, and all methods, with the exception of the non-optimized Lattice Boltzmann method, are run with the -O3 flag active. We ran each system for 1000 iterations, and initialized each system by utilizing the same methods as in section 4.}
    \label{fig:timing}
\end{figure}

To compare the performance of the different algorithms we used the simulations of a decaying sign wave presented in Sec. (4) for varying average numbers of particles. The setup consists of a 32x32 lattice with an initial density distribution given by Eq. (\ref{eqn:Prho}) with an initial amplitude of $A=N^{eq}$, as indicated in Eq. (\ref{Ath}). Since the lookup table method generates sample points as the simulation runs, we first run the simulation for 1000 iterations to ensure that the sampling algorithm has fully initialized and then measure the runtime of the next 1000 iterations.

The results for an inverse relaxation time of $\omega=1$ are shown in Figure \ref{fig:timing}. For the Collision method we cannot guarantee that all particles will have collided, and therefore we chose a number of collisions that ensures that on average 99.9\% of particles will have collided, as explained in Appendix \ref{AppendixB}. As expected, the number of collisions scales linearly with the number or particles, as opposed to quadratically with the number of particles in the case of binary collisions in the hydrodynamic lattice gas \cite{blommel2018integer}. The lattice gas implementation with the sampling collision operator performs significantly better than the collision algorithm. It even outperforms the non-optimized version of the lattice Boltzmann code, although it remains slower than the optimized lattice Boltzmann code by about a factor of five. We believe that there is still significant potential for optimization, not least in the sampling algorithm. The implementation of the lookup method for sampling binomial numbers already outperforms the GSL algorithm in the range between 20 and 2000 average numbers of particles per lattice site. It is remarkable that the GSL sampling algorithms speed up for average numbers of particles larger than 2000, which will allow the algorithm to scale well for large numbers of particles. 
\begin{figure}
    \centering
    \subfloat[]{
    \includegraphics[width=0.9\columnwidth]{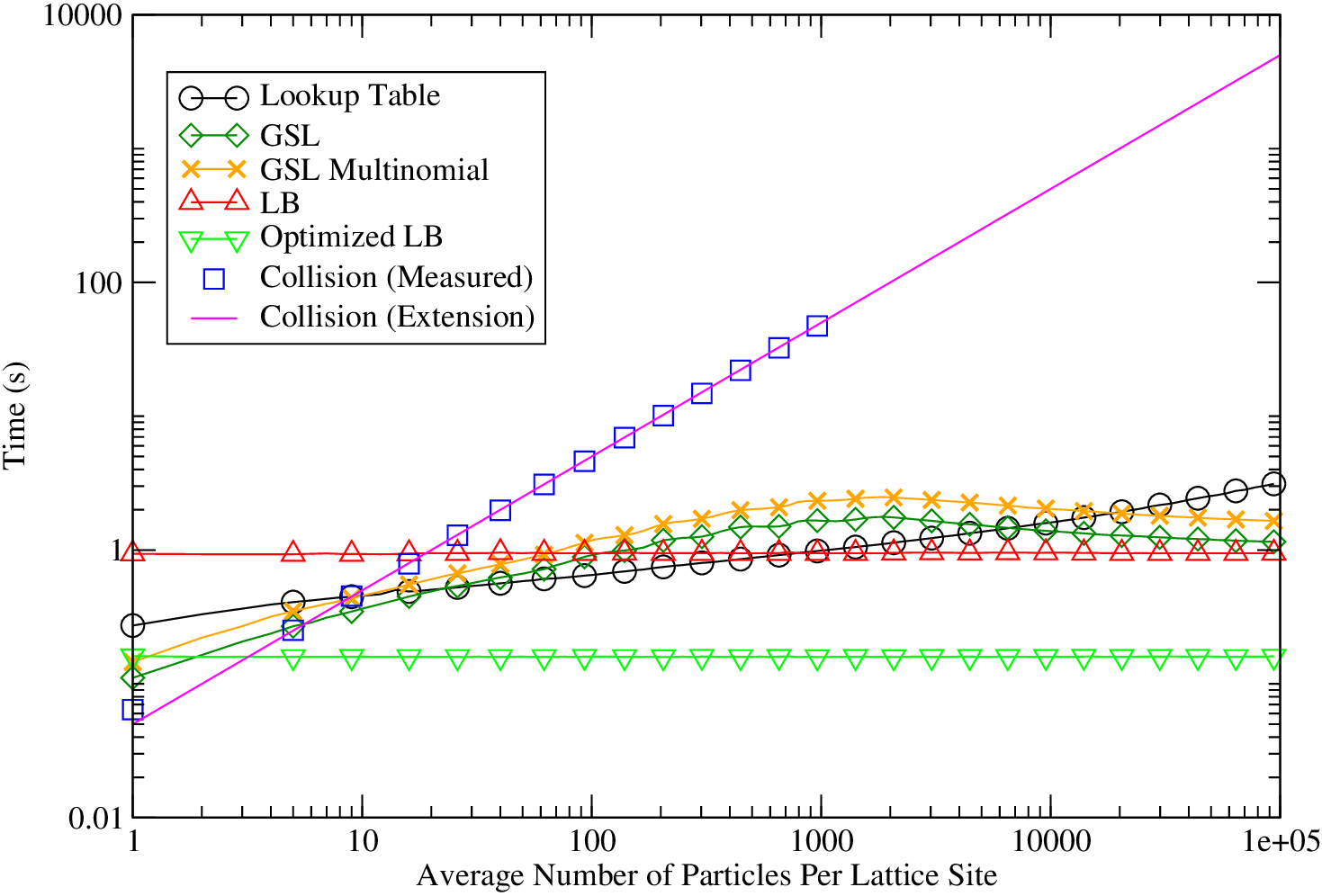}}\\
    \subfloat[]{
    \includegraphics[width=0.9\columnwidth]{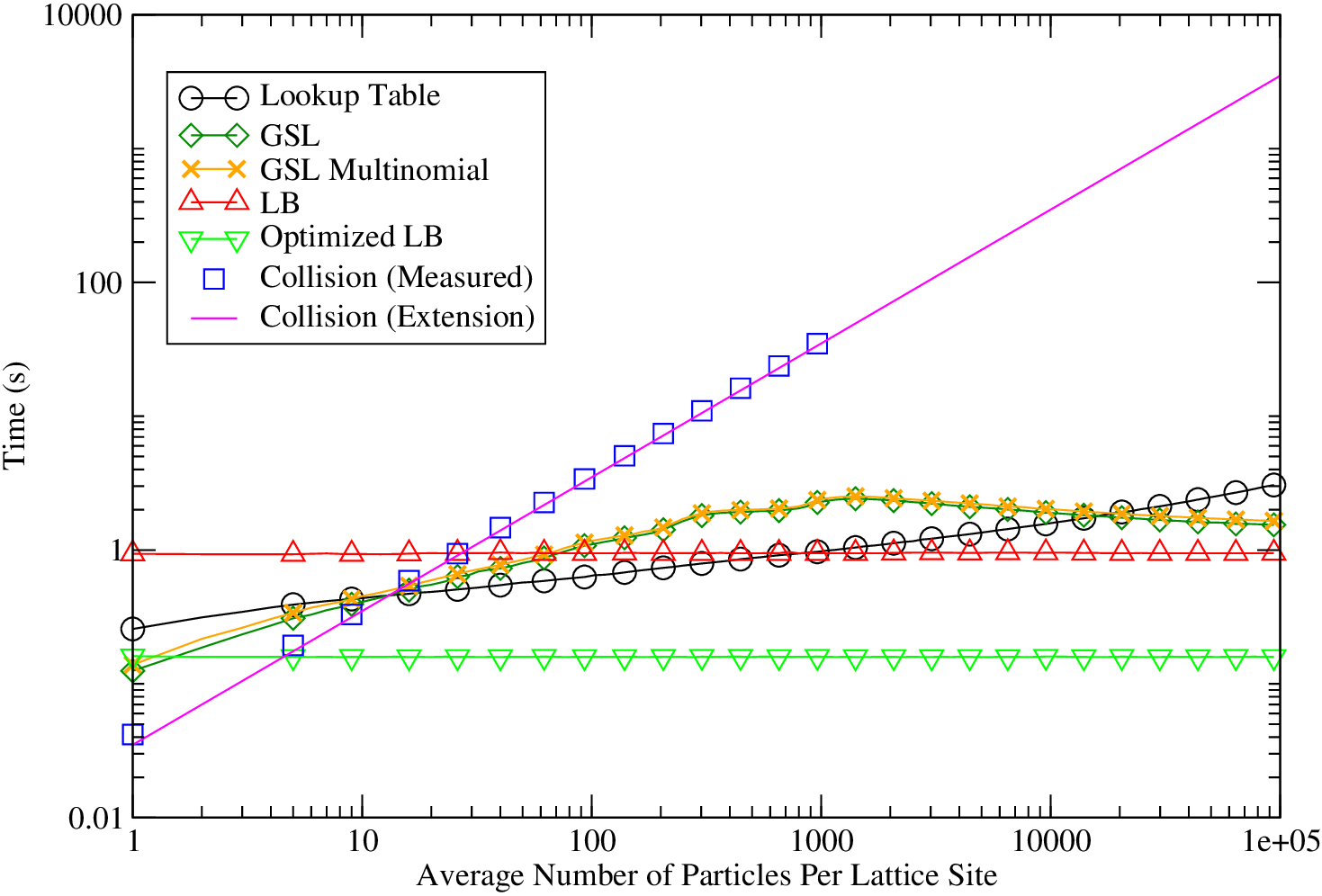}}\\
    \subfloat[]{
    \includegraphics[width=0.9\columnwidth]{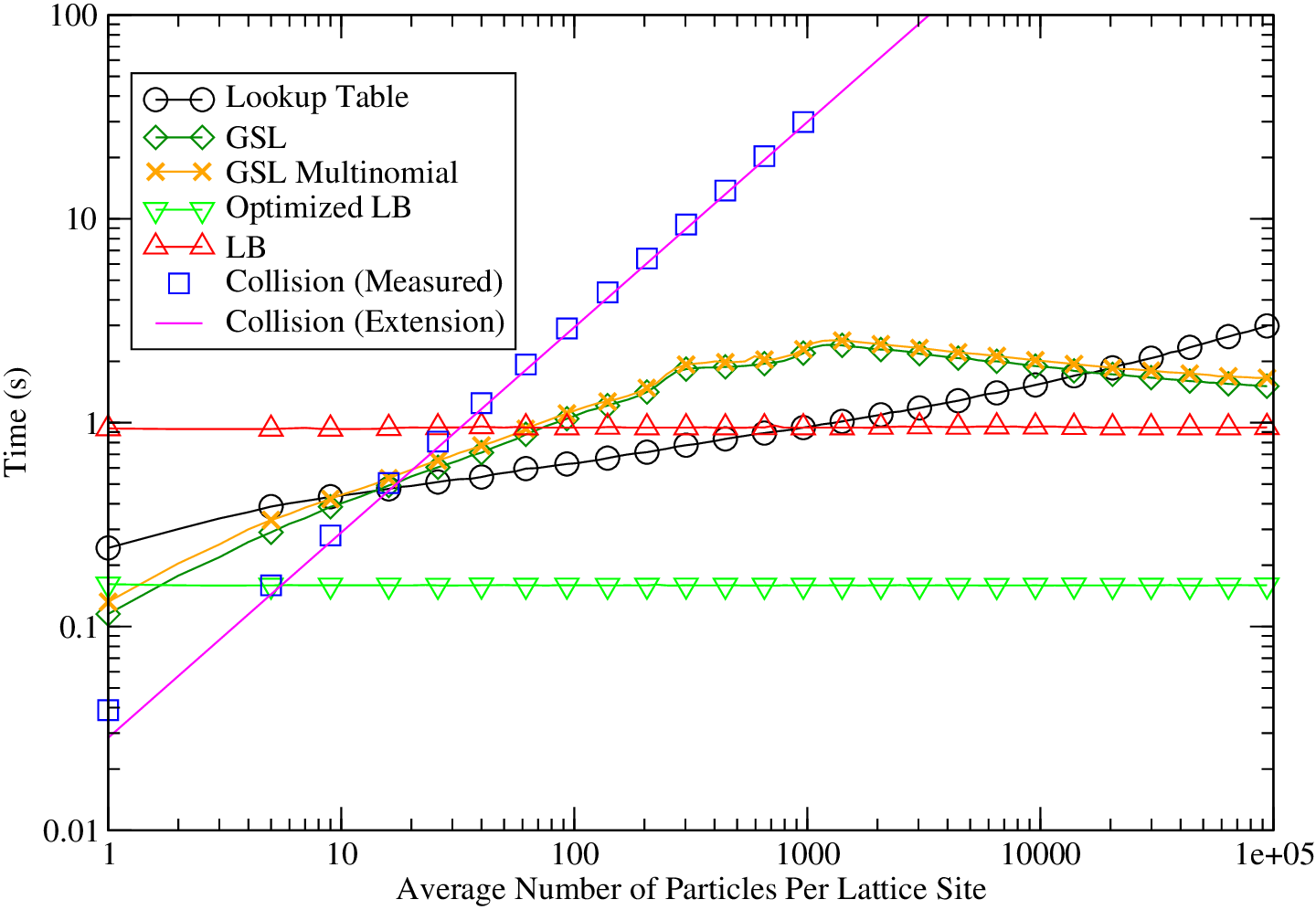}}
    \caption{Execution time of the different versions of the sampling algorithm compared to the original integer lattice gas approach and two lattice Boltzmann implementations, with the relaxation time $
    1/\omega$ being 1.5 in \ref{fig:timingtau}a, 1.8 in \ref{fig:timingtau}b, and 2.0 in \ref{fig:timingtau}c graphed on a log-log scale. Each method was run utilizing the same parameters as the previously shown timing graph, including the grid size, number of iterations, and initialization method. The extension of the collision method is an approximate line of best fit for the measured collision data, which shows how the data would progress at higher numbers of particles.}
    \label{fig:timingtau}
\end{figure}

As we decrease the inverse collision time $\omega$ we have to perform fewer collisions, which benefits the collision method. The effect on the sampling method is harder to predict: on the one hand a second sampling step is required to select the particles to be collided, as indicated in Eq. (\ref{eqn:Nselect}); on the other hand the number of particles to be collided are fewer, which generally leads to faster sampling times. The net effect of this is shown in Figure \ref{fig:timingtau}. As expected the collision method becomes faster with increasing $\omega$. The timing of the sampling methods changes only slightly. The most notable change is for the GSL methods, which show a slight increase in the execution time for the sequential multinomial sampling algorithm, and a significant increase in the execution time for the direct multinomial sampling algorithm, to the point that, while the direct sampling algorithm is faster for $\omega=1$, it is slower for $\omega<1$. We are not sure what the reason for this unexpected behavior of the library calls is.  However, even for $\omega=0.5$ the sampling collision operators continue to outperform the direct collision approach when there are more than 20 particles per lattice site on average.  Thus, for most practical approaches the Monte Carlo lattice gas is superior to the collision method, and has the potential to become competitive  with the fluctuating lattice Boltzmann method.

\section{Conclusions}
We developed a novel integer lattice gas method for the fluctuating diffusion equation. This method remedies some deficiencies of the equivalent fluctuating lattice Boltzmann method, particularly for small densities. The fundamental approach based on single particle collisions and equivalent to the approach pioneered by Blommel for hydrodynamics integer lattice gases \cite{blommel2018integer} becomes slow for larger numbers of particles per lattice site. The computational cost for this approach scales linearly with the number of particles. To remedy this difficulty we also developed a sampling collision operator that picks new particle distributions directly from a local equilibrium distribution. The runtime of this algorithm now scales much better, and, in the case of the GSL sampling algorithm, approximately recovers the flat scaling of lattice Boltzmann approaches at higher densities. We believe that this first algorithm of a sampling collision operator shows the potential for lattice gas approaches to become competitive with lattice Boltzmann approaches again. These results open the door to develop sampling collision operators for hydrodynamic integer lattice gases that we hope will be computationally competitive with fluctuating lattice Boltzmann methods and have correct statistics for small numbers of particles per lattice site. 

\appendix
\section{Efficient Binomially distributed random numbers\label{code}}

There are quite a few efficient Binomial sampling algorithms available, such as the GNU Scientific Library (GSL) which is the algorithm used for large densities on this project. We also developed our own algorithm, that turns out to be faster for some range of parameters. This is encouraging since a key cost of the lattice gas algorithm with a sampling collision operator lies in the generation of binomially distributed random numbers. A second motivation for developing an in-house sampling algorithm lies in the fact that hydrodynamic sampling algorithms we need to sample from other discrete (and as yet unnamed) distributions, and our algorithm easily generalizes for this case. Better algorithm implementations are also likely to be possible, which would lead to a much faster simulation\cite{Agrawal2018,pmlr-v108-saad20a}.

\begin{figure}
    \centering
    \subfloat[]{
    \includegraphics[width=\columnwidth]{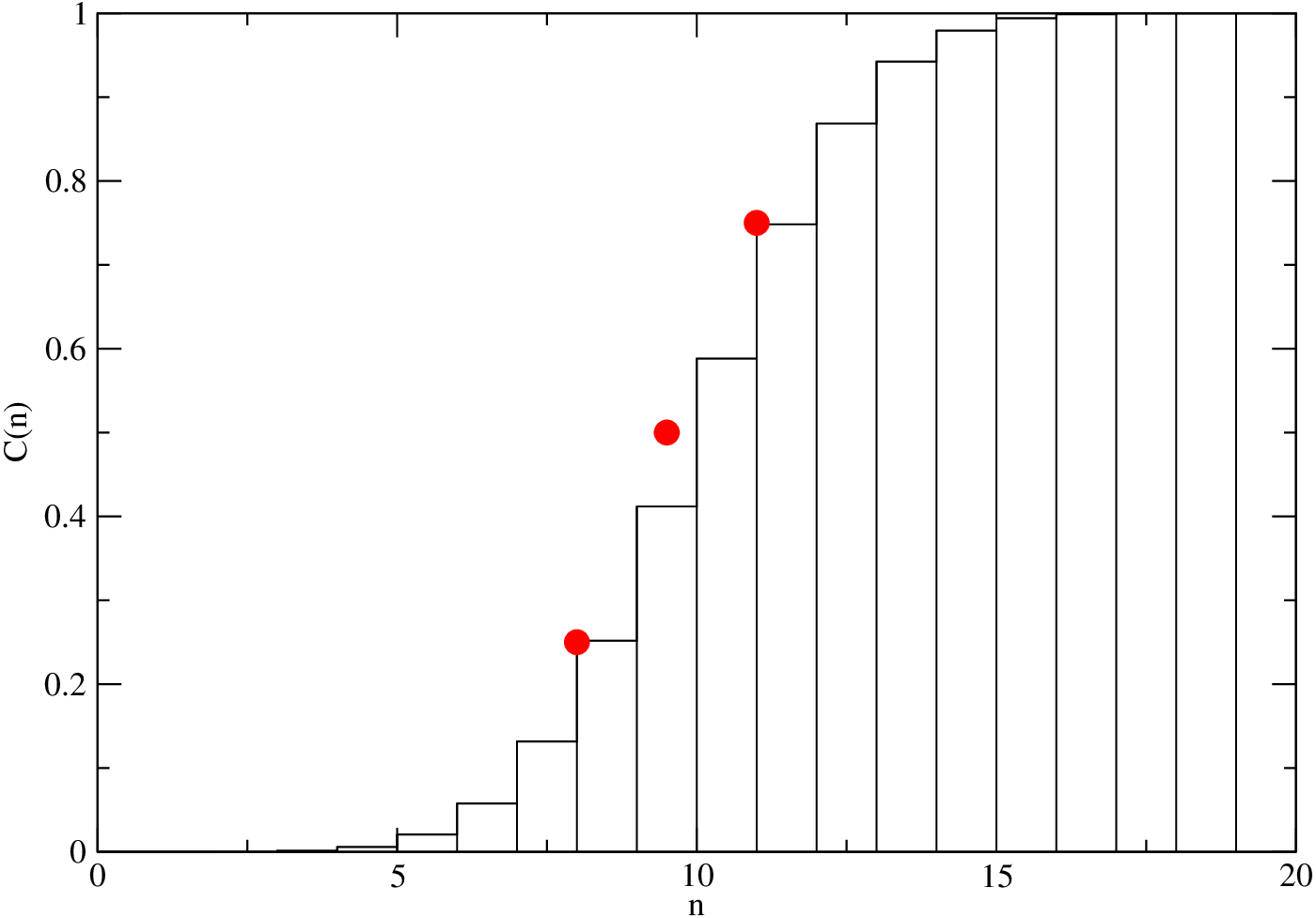}}\\
     \subfloat[]{
    \includegraphics[width=\columnwidth]{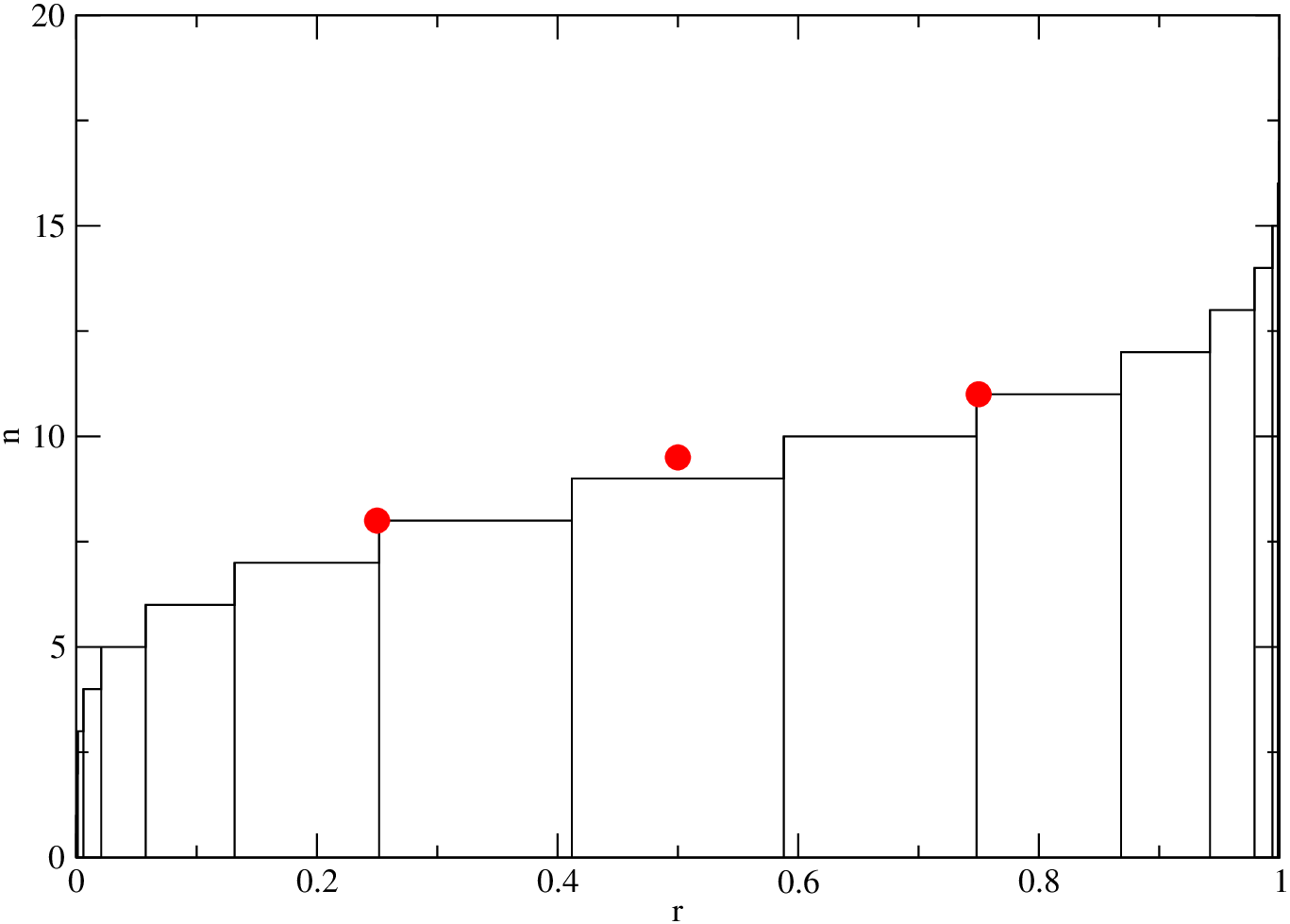}}
    \caption{Basic idea of generating random numbers according to a prescribed probability distribution function with $N=20$ and $p=0.5$. In (a) we obtain the cumulative distribution function, and in (b) we invert the cumulative distribution function, and  use it to obtain the appropriately distributed integer random numbers from Eq. (\ref{eqn:MC}). Each dot represents a number saved by the lookup table. If the point and the graph meet, the point's number is saved directly, if not, then whatever the last point checked that was on the graph becomes the saved point.}
    \label{fig:sample}
\end{figure}

Our algorithm follows from the concept of inverting a cumulative distribution function, as explained e.g. in the Numerical Recipes \cite[chapter 7.3.2]{press2007numerical}. Unfortunately, there is no explicit expression for the cumulative distribution function $C(n)$, instead it is possible to derive a recursion relation that gives $C(n)$ through
\begin{equation}
    C(n)=\sum_{m=0}^n P(m).
\end{equation}
Given a random number $0\leq r\leq 1$ we find a suitably distributed integer random number $M$, given by:
\begin{equation}
    M=C^{-1}(r)
    \label{eqn:MC}
\end{equation}
as illustrated in Figure \ref{fig:sample}.


For the binomial distribution we have an analytical expression
\begin{equation}
    P(n) = \frac{N!}{n!((N-n)!}p^n (1-p)^{N-n}
\end{equation}
but for larger $N$ the evaluation of the factorial as well as the power of $p$ can cause potential  difficulties. 

However, we do not have an explicit formula for calculating the cumulative distribution function. We can generate it through the recursive relations
\begin{align}
     P(0)&=(1-p)^N\nonumber,\\
     P(n+1)&=P(n)\frac{p(N-n+1)}{(1-p)n}\nonumber,\\
     P(n)&=P(n+1)\frac{(1-p)n}{p(N-n+1)}.
     \label{eqn:Crec}
 \end{align}
 Since $C(n)$ is the sum of $P(n)$, these relations allow us to generate the cumulative distribution function as needed. For $N\gtrapprox 250$, there is a problem of numerical underflow. To mitigate this issue, we began to utilize the logarithms of these recursion relations
 \begin{align}
     \ln[P(0)]=&N\ln(1-p)\nonumber\\
     \ln[P(n+1)]=&\ln[P(n)]\nonumber\\&+\ln[p(N-n+1)]-\ln[(1-p)n]\nonumber\\
     \ln[P(n)]=&\ln[P(n+1)]\nonumber\\&+\ln[(1-p)n]-\ln[p(N-n+1)].
     \label{eqn:Clnrec}
 \end{align}
 Thus, the algorithm consists of generating a uniformly distributed random number $r$ between 0 and 1, and then find a binomially distributed random number $n$ by the condition 
 \begin{equation}
     C(n-1)\leq r<C(n)
 \end{equation}
 where we defined $C(-1)=0$. 
 This approach has two key drawbacks: firstly, using logarithms for calculating $C(n)$ iteratively is computationally very expensive; secondly, for large $N$ the recursive algorithm will take on average $N/2$ iterations to obtain the required $n$.
 
 To alleviate this problem we devised a lookup method, where we pre-calculate the values of $n$ for $r=0.25$, $r=0.5$ and $r=0.75$ and store them in memory. This allows a more targeted search which both allows using Eq. (\ref{eqn:Crec}) instead of Eq. (\ref{eqn:Clnrec}) to obtain $C(n)$ from a closer starting point to the target value. The number of iterations required is also substantially reduced, which is shown in Figure \ref{fig:sample}. We also experimented with using more saved values, but the required overhead essentially outweighed the gains for the code we generated.

 \section{Determining the Equivalent Number if Collisions Per Timestep to match specific values of $\omega$\label{AppendixB}}
In order to compare the execution times of the Collision lattice gas method and the Monte Carlo lattice gas method, it was necessary to relate the values of $\omega$ used for sampling the multinomial distribution to the number of collisions per lattice site per timestep. In order to do this, we made an estimation that collisions within each lattice site are independent, and looked at the average fraction of particles that remains uncollided, which is the definition of the value of $\omega$. Thus, taking the number of uncollided particles $N_u$ out of a system of $N$ particles as a continuous function of collision time $c$, we can approximate the discrete collision process by a continuous differential equation:
\begin{equation}
    \frac{d}{dc}N_u(c)=-\frac{N_u(c)}{N},
\end{equation}
where $N_u(c)/N$ is the probability that a randomly picked particle has not undergone a collision before.
With the initial condition $N_u(0)=N$ this gives
\begin{equation}
    N_u(c)=N\exp\left(-\frac{c}{N}\right).
\end{equation}
From the lattice Boltzmann evolution equation (\ref{MCLB}) with the BGK collision operator of Eq. (\ref{eqn:LBcoll}) we see that the fraction of uncollided particles is given by $1-\omega$. We therefore get
\begin{equation}
\frac{N_u(c)}{N}=1-\omega.
\end{equation}
We then take this equation, and, substituting the total number of collisions $C$ for the collision time $c$, we obtain:
\begin{equation}
    1-\omega=\exp\left(-\frac{C}{N}\right).
\end{equation}
Solving for $C$ gives:
\begin{equation}
    C=-N\ln(1-\omega),
\end{equation}
which is the relation for the number of collisions per timestep for a number of particles $N$ and a given $\omega$.

\bibliography{AW,MCLG,IntegerLG,LB}

\begin{thebibliography}{17}%
\makeatletter
\providecommand \@ifxundefined [1]{%
 \@ifx{#1\undefined}
}%
\providecommand \@ifnum [1]{%
 \ifnum #1\expandafter \@firstoftwo
 \else \expandafter \@secondoftwo
 \fi
}%
\providecommand \@ifx [1]{%
 \ifx #1\expandafter \@firstoftwo
 \else \expandafter \@secondoftwo
 \fi
}%
\providecommand \natexlab [1]{#1}%
\providecommand \enquote  [1]{``#1''}%
\providecommand \bibnamefont  [1]{#1}%
\providecommand \bibfnamefont [1]{#1}%
\providecommand \citenamefont [1]{#1}%
\providecommand \href@noop [0]{\@secondoftwo}%
\providecommand \href [0]{\begingroup \@sanitize@url \@href}%
\providecommand \@href[1]{\@@startlink{#1}\@@href}%
\providecommand \@@href[1]{\endgroup#1\@@endlink}%
\providecommand \@sanitize@url [0]{\catcode `\\12\catcode `\$12\catcode
  `\&12\catcode `\#12\catcode `\^12\catcode `\_12\catcode `\%12\relax}%
\providecommand \@@startlink[1]{}%
\providecommand \@@endlink[0]{}%
\providecommand \url  [0]{\begingroup\@sanitize@url \@url }%
\providecommand \@url [1]{\endgroup\@href {#1}{\urlprefix }}%
\providecommand \urlprefix  [0]{URL }%
\providecommand \Eprint [0]{\href }%
\providecommand \doibase [0]{http://dx.doi.org/}%
\providecommand \selectlanguage [0]{\@gobble}%
\providecommand \bibinfo  [0]{\@secondoftwo}%
\providecommand \bibfield  [0]{\@secondoftwo}%
\providecommand \translation [1]{[#1]}%
\providecommand \BibitemOpen [0]{}%
\providecommand \bibitemStop [0]{}%
\providecommand \bibitemNoStop [0]{.\EOS\space}%
\providecommand \EOS [0]{\spacefactor3000\relax}%
\providecommand \BibitemShut  [1]{\csname bibitem#1\endcsname}%
\let\auto@bib@innerbib\@empty
\bibitem [{\citenamefont {Blommel}\ and\ \citenamefont
  {Wagner}(2018)}]{blommel2018integer}%
  \BibitemOpen
  \bibfield  {author} {\bibinfo {author} {\bibfnamefont {Thomas}\ \bibnamefont
  {Blommel}}\ and\ \bibinfo {author} {\bibfnamefont {Alexander~J}\ \bibnamefont
  {Wagner}},\ }\bibfield  {title} {\enquote {\bibinfo {title} {Integer lattice
  gas with monte carlo collision operator recovers the lattice boltzmann method
  with poisson-distributed fluctuations},}\ }\href@noop {} {\bibfield
  {journal} {\bibinfo  {journal} {Physical Review E}\ }\textbf {\bibinfo
  {volume} {97}},\ \bibinfo {pages} {023310} (\bibinfo {year}
  {2018})}\BibitemShut {NoStop}%
\bibitem [{\citenamefont {Frisch}\ \emph {et~al.}(1986)\citenamefont {Frisch},
  \citenamefont {Hasslacher},\ and\ \citenamefont {Pomeau}}]{FHP}%
  \BibitemOpen
  \bibfield  {author} {\bibinfo {author} {\bibfnamefont {Uriel}\ \bibnamefont
  {Frisch}}, \bibinfo {author} {\bibfnamefont {Brosl}\ \bibnamefont
  {Hasslacher}}, \ and\ \bibinfo {author} {\bibfnamefont {Yves}\ \bibnamefont
  {Pomeau}},\ }\bibfield  {title} {\enquote {\bibinfo {title} {Lattice-gas
  automata for the navier-stokes equation},}\ }\href@noop {} {\bibfield
  {journal} {\bibinfo  {journal} {Physical Review Letters}\ }\textbf {\bibinfo
  {volume} {56}},\ \bibinfo {pages} {1505} (\bibinfo {year}
  {1986})}\BibitemShut {NoStop}%
\bibitem [{\citenamefont {Doolen}(1991)}]{doolen1991lattice}%
  \BibitemOpen
  \bibfield  {author} {\bibinfo {author} {\bibfnamefont {Gary~D}\ \bibnamefont
  {Doolen}},\ }\href@noop {} {\emph {\bibinfo {title} {Lattice gas methods:
  theory, applications, and hardware}}},\ Vol.~\bibinfo {volume} {47}\
  (\bibinfo  {publisher} {MIT press},\ \bibinfo {year} {1991})\BibitemShut
  {NoStop}%
\bibitem [{\citenamefont {Wolf-Gladrow}(2004)}]{wolf2004lattice}%
  \BibitemOpen
  \bibfield  {author} {\bibinfo {author} {\bibfnamefont {Dieter~A}\
  \bibnamefont {Wolf-Gladrow}},\ }\href@noop {} {\emph {\bibinfo {title}
  {Lattice-gas cellular automata and lattice Boltzmann models: an
  introduction}}}\ (\bibinfo  {publisher} {Springer},\ \bibinfo {year}
  {2004})\BibitemShut {NoStop}%
\bibitem [{\citenamefont {Higuera}\ and\ \citenamefont
  {Jimenez}(1989)}]{higuera1989boltzmann}%
  \BibitemOpen
  \bibfield  {author} {\bibinfo {author} {\bibfnamefont {FJ}~\bibnamefont
  {Higuera}}\ and\ \bibinfo {author} {\bibfnamefont {J}~\bibnamefont
  {Jimenez}},\ }\bibfield  {title} {\enquote {\bibinfo {title} {Boltzmann
  approach to lattice gas simulations},}\ }\href@noop {} {\bibfield  {journal}
  {\bibinfo  {journal} {EPL (Europhysics Letters)}\ }\textbf {\bibinfo {volume}
  {9}},\ \bibinfo {pages} {663} (\bibinfo {year} {1989})}\BibitemShut {NoStop}%
\bibitem [{\citenamefont {Qian}\ \emph {et~al.}(1992)\citenamefont {Qian},
  \citenamefont {d'Humi{\`e}res},\ and\ \citenamefont
  {Lallemand}}]{qian1992lattice}%
  \BibitemOpen
  \bibfield  {author} {\bibinfo {author} {\bibfnamefont {YH}~\bibnamefont
  {Qian}}, \bibinfo {author} {\bibfnamefont {Dominique}\ \bibnamefont
  {d'Humi{\`e}res}}, \ and\ \bibinfo {author} {\bibfnamefont {Pierre}\
  \bibnamefont {Lallemand}},\ }\bibfield  {title} {\enquote {\bibinfo {title}
  {Lattice bgk models for navier-stokes equation},}\ }\href@noop {} {\bibfield
  {journal} {\bibinfo  {journal} {EPL (Europhysics Letters)}\ }\textbf
  {\bibinfo {volume} {17}},\ \bibinfo {pages} {479} (\bibinfo {year}
  {1992})}\BibitemShut {NoStop}%
\bibitem [{\citenamefont {Boghosian}\ \emph {et~al.}(1997)\citenamefont
  {Boghosian}, \citenamefont {Yepez}, \citenamefont {Alexander},\ and\
  \citenamefont {Margolus}}]{boghosian1997integer}%
  \BibitemOpen
  \bibfield  {author} {\bibinfo {author} {\bibfnamefont {Bruce~M}\ \bibnamefont
  {Boghosian}}, \bibinfo {author} {\bibfnamefont {Jeffrey}\ \bibnamefont
  {Yepez}}, \bibinfo {author} {\bibfnamefont {Francis~J}\ \bibnamefont
  {Alexander}}, \ and\ \bibinfo {author} {\bibfnamefont {Norman~H}\
  \bibnamefont {Margolus}},\ }\bibfield  {title} {\enquote {\bibinfo {title}
  {Integer lattice gases},}\ }\href@noop {} {\bibfield  {journal} {\bibinfo
  {journal} {Physical Review E}\ }\textbf {\bibinfo {volume} {55}},\ \bibinfo
  {pages} {4137} (\bibinfo {year} {1997})}\BibitemShut {NoStop}%
\bibitem [{\citenamefont {Wolf-Gladrow}(1995)}]{wolf1995lattice}%
  \BibitemOpen
  \bibfield  {author} {\bibinfo {author} {\bibfnamefont {Dieter}\ \bibnamefont
  {Wolf-Gladrow}},\ }\bibfield  {title} {\enquote {\bibinfo {title} {A lattice
  boltzmann equation for diffusion},}\ }\href@noop {} {\bibfield  {journal}
  {\bibinfo  {journal} {Journal of statistical physics}\ }\textbf {\bibinfo
  {volume} {79}},\ \bibinfo {pages} {1023--1032} (\bibinfo {year}
  {1995})}\BibitemShut {NoStop}%
\bibitem [{\citenamefont {Wagner}\ and\ \citenamefont
  {Strand}(2016)}]{wagner2016fluctuating}%
  \BibitemOpen
  \bibfield  {author} {\bibinfo {author} {\bibfnamefont {Alexander~J}\
  \bibnamefont {Wagner}}\ and\ \bibinfo {author} {\bibfnamefont {Kyle}\
  \bibnamefont {Strand}},\ }\bibfield  {title} {\enquote {\bibinfo {title}
  {Fluctuating lattice boltzmann method for the diffusion equation},}\
  }\href@noop {} {\bibfield  {journal} {\bibinfo  {journal} {Physical Review
  E}\ }\textbf {\bibinfo {volume} {94}},\ \bibinfo {pages} {033302} (\bibinfo
  {year} {2016})}\BibitemShut {NoStop}%
\bibitem [{\citenamefont {Guichard}(2021)}]{Guichard}%
  \BibitemOpen
  \bibfield  {author} {\bibinfo {author} {\bibfnamefont {David}\ \bibnamefont
  {Guichard}},\ }\href@noop {} {\emph {\bibinfo {title} {Combinatorics and
  Graph Theory}}}\ (\bibinfo {year} {2021})\ \bibinfo {note}
  {\url{http://www.whitman.edu/mathematics/cgt_online/cgt.pdf}}\BibitemShut
  {NoStop}%
\bibitem [{\citenamefont {Galassi}(2018)}]{GNU}%
  \BibitemOpen
  \bibfield  {author} {\bibinfo {author} {\bibfnamefont {M.~et~al}\
  \bibnamefont {Galassi}},\ }\href {https://www.gnu.org/software/gsl/}
  {\enquote {\bibinfo {title} {Gnu scientific library reference manual},}\ }
  (\bibinfo {year} {2018})\BibitemShut {NoStop}%
\bibitem [{\citenamefont {Landau}\ and\ \citenamefont
  {Lifshitz}(1969)}]{landau1969statistical}%
  \BibitemOpen
  \bibfield  {author} {\bibinfo {author} {\bibfnamefont {Lev~Davidovich}\
  \bibnamefont {Landau}}\ and\ \bibinfo {author} {\bibfnamefont
  {EM}~\bibnamefont {Lifshitz}},\ }\bibfield  {title} {\enquote {\bibinfo
  {title} {Statistical physics. pt. 1},}\ }\href@noop {} {\bibfield  {journal}
  {\bibinfo  {journal} {Course of theoretical physics-Pergamon International
  Library of Science, Technology, Engineering and Social Studies, Oxford:
  Pergamon Press, and Reading: Addison-Wesley,| c1969, 2nd rev.-enlarg. ed.}\ }
  (\bibinfo {year} {1969})}\BibitemShut {NoStop}%
\bibitem [{\citenamefont {Gough}(2009)}]{gough2009gnu}%
  \BibitemOpen
  \bibfield  {author} {\bibinfo {author} {\bibfnamefont {Brian}\ \bibnamefont
  {Gough}},\ }\href@noop {} {\emph {\bibinfo {title} {GNU scientific library
  reference manual}}}\ (\bibinfo  {publisher} {Network Theory Ltd.},\ \bibinfo
  {year} {2009})\BibitemShut {NoStop}%
\bibitem [{\citenamefont {Seekins}(2021)}]{SeekinsDiffMCLG2021Git}%
  \BibitemOpen
  \bibfield  {author} {\bibinfo {author} {\bibfnamefont {Noah}\ \bibnamefont
  {Seekins}},\ }\href@noop {} {\enquote {\bibinfo {title} {Monte carlo lattice
  gas},}\ }\bibinfo {howpublished}
  {\url{https://github.com/NoahSeekins/MonteCarloLatticeGas}} (\bibinfo {year}
  {2021})\BibitemShut {NoStop}%
\bibitem [{\citenamefont {Agrawal}\ \emph {et~al.}(2018)\citenamefont
  {Agrawal}, \citenamefont {Bhattacharya},\ and\ \citenamefont
  {Ansumali}}]{Agrawal2018}%
  \BibitemOpen
  \bibfield  {author} {\bibinfo {author} {\bibfnamefont {Samarth}\ \bibnamefont
  {Agrawal}}, \bibinfo {author} {\bibfnamefont {Soumyadeep}\ \bibnamefont
  {Bhattacharya}}, \ and\ \bibinfo {author} {\bibfnamefont {Santosh}\
  \bibnamefont {Ansumali}},\ }\bibfield  {title} {\enquote {\bibinfo {title}
  {Molecular dice: Random number generators \'a la boltzmann},}\ }\href
  {\doibase 10.1103/PhysRevE.98.063315} {\bibfield  {journal} {\bibinfo
  {journal} {Phys. Rev. E}\ }\textbf {\bibinfo {volume} {98}},\ \bibinfo
  {pages} {063315} (\bibinfo {year} {2018})}\BibitemShut {NoStop}%
\bibitem [{\citenamefont {Saad}\ \emph {et~al.}(2020)\citenamefont {Saad},
  \citenamefont {Freer}, \citenamefont {Rinard},\ and\ \citenamefont
  {Mansinghka}}]{pmlr-v108-saad20a}%
  \BibitemOpen
  \bibfield  {author} {\bibinfo {author} {\bibfnamefont {Feras}\ \bibnamefont
  {Saad}}, \bibinfo {author} {\bibfnamefont {Cameron}\ \bibnamefont {Freer}},
  \bibinfo {author} {\bibfnamefont {Martin}\ \bibnamefont {Rinard}}, \ and\
  \bibinfo {author} {\bibfnamefont {Vikash}\ \bibnamefont {Mansinghka}},\
  }\bibfield  {title} {\enquote {\bibinfo {title} {The fast loaded dice roller:
  A near-optimal exact sampler for discrete probability distributions},}\ }in\
  \href {http://proceedings.mlr.press/v108/saad20a.html} {\emph {\bibinfo
  {booktitle} {Proceedings of the Twenty Third International Conference on
  Artificial Intelligence and Statistics}}},\ \bibinfo {series} {Proceedings of
  Machine Learning Research}, Vol.\ \bibinfo {volume} {108},\ \bibinfo {editor}
  {edited by\ \bibinfo {editor} {\bibfnamefont {Silvia}\ \bibnamefont
  {Chiappa}}\ and\ \bibinfo {editor} {\bibfnamefont {Roberto}\ \bibnamefont
  {Calandra}}}\ (\bibinfo  {publisher} {PMLR},\ \bibinfo {year} {2020})\ pp.\
  \bibinfo {pages} {1036--1046}\BibitemShut {NoStop}%
\bibitem [{\citenamefont {Press}\ \emph {et~al.}(2007)\citenamefont {Press},
  \citenamefont {William}, \citenamefont {Teukolsky}, \citenamefont {Saul},
  \citenamefont {Vetterling},\ and\ \citenamefont
  {Flannery}}]{press2007numerical}%
  \BibitemOpen
  \bibfield  {author} {\bibinfo {author} {\bibfnamefont {William~H}\
  \bibnamefont {Press}}, \bibinfo {author} {\bibfnamefont {H}~\bibnamefont
  {William}}, \bibinfo {author} {\bibfnamefont {Saul~A}\ \bibnamefont
  {Teukolsky}}, \bibinfo {author} {\bibfnamefont {A}~\bibnamefont {Saul}},
  \bibinfo {author} {\bibfnamefont {William~T}\ \bibnamefont {Vetterling}}, \
  and\ \bibinfo {author} {\bibfnamefont {Brian~P}\ \bibnamefont {Flannery}},\
  }\href@noop {} {\emph {\bibinfo {title} {Numerical recipes 3rd edition: The
  art of scientific computing}}}\ (\bibinfo  {publisher} {Cambridge university
  press},\ \bibinfo {year} {2007})\BibitemShut {NoStop}%
\end{thebibliography}%

\end{document}